\documentclass[%
 superscriptaddress,
 amsmath,amssymb,
 aps,
 prb,
 twocolumn,
 longbibliography %, linenumbers
]{revtex4-2}

\usepackage{mathtools} %Extension of amsmath. Better matrices
\usepackage{graphicx}% Include figure files
\usepackage{dcolumn}% Align table columns on decimal point
\usepackage{bm}% bold math
\usepackage{xcolor}
\usepackage{tikz}
\usepackage{xfrac}
\usepackage{blindtext}
\usepackage{times}
\usepackage{multirow}
\usepackage{chemformula}
\usepackage{float}
\usepackage{comment}
\usepackage{ragged2e}
\usepackage{placeins} %FloatBarrier command 

\usepackage{color}
\usepackage[colorlinks=true, urlcolor=blue, linkcolor=blue, citecolor=blue, pdftex]{hyperref}

\usepackage[normalem]{ulem}

\definecolor{darkred}{rgb}{0.55, 0.0, 0.0}
\definecolor{darkblue}{rgb}{0.0, 0.0, 0.55}
\usepackage{hyperref}
\hypersetup{
	colorlinks=true,
	urlcolor=darkred,
	citecolor=darkblue,
	linkcolor=darkred,
	breaklinks
}

\graphicspath{{figures/}}

\newcommand{\be}{\begin{equation}}
\newcommand{\ee}{\end{equation}}

\newcommand{\bea}{\begin{eqnarray}}
\newcommand{\eea}{\end{eqnarray}}
\newcommand{\beal}{\begin{align}}
\newcommand{\eeal}{\end{align}}

\newcommand{\goto}{\rightarrow}

\renewcommand{\vec}[1]{\ensuremath{\mathbf{#1}}}

\newcommand{\abs}[1]{\left| #1 \right|} % for absolute value
\newcommand{\avg}[1]{\left< #1 \right>} % for average
\newcommand{\GP}{\Gamma^\prime}
\newcommand{\pr}{^{\prime}}

\begin{document}

\title{Constructing Emergent U(1) Symmetries in the Gamma-prime $\left(\bf \GP\right)$ model}

\author{Sagar Ramchandani}
\affiliation{Institute for Theoretical Physics, University of Cologne, 50937 Cologne, Germany.}

\author{Simon Trebst}
\affiliation{Institute for Theoretical Physics, University of Cologne, 50937 Cologne, Germany.}

\author{Ciar\'an Hickey}
\affiliation{Institute for Theoretical Physics, University of Cologne, 50937 Cologne, Germany.}
\affiliation{School of Physics, University College Dublin, Belfield, Dublin 4, Ireland.}
\affiliation{Centre for Quantum Engineering, Science, and Technology, University College Dublin, Dublin 4, Ireland}

\begin{abstract}
    Frustrated magnets can elude the paradigm of conventional symmetry breaking and instead exhibit signatures of {\em emergent} symmetries at low temperatures. Such symmetries arise from ``accidental" degeneracies within the ground state manifold and have been explored in a number of disparate models, in both two and three dimensions. 
    Here we report the systematic construction of a family of classical spin models that, for a wide variety of lattice geometries with triangular motifs
    in one, two and three spatial dimensions, such as the kagome or hyperkagome lattices, exhibit an emergent, continuous U(1) symmetry.
    This is particularly surprising because the underlying Hamiltonian actually has very little symmetry
    — a bond-directional, off-diagonal exchange model inspired by the microscopics of spin-orbit entangled materials (the $\GP$-model).
    The construction thus allows for a systematic study of the interplay between the emergent continuous U(1) symmetry and the underlying discrete Hamiltonian symmetries in different lattices across different spatial dimensions.
    We discuss the impact of thermal and quantum fluctuations in lifting the accidental ground state degeneracy via the thermal and quantum order-by-disorder mechanisms, and how spatial dimensionality and lattice symmetries play a crucial role in shaping the physics of the model. Complementary Monte Carlo simulations, for representative one-, two-, and three-dimensional lattice geometries, provide a complete account of the thermodynamics and confirm our analytical expectations.  
\end{abstract}
\maketitle

%%%%%%%%%%%%%%%%%%%%%%%%%%%%%%%%%%%%%%%%%%%%
%% Introduction
%%%%%%%%%%%%%%%%%%%%%%%%%%%%%%%%%%%%%%%%%%%%

Magnetic frustration can induce a wide range of fascinating phenomena beyond conventional magnetic order,
including the formation of
valence bond crystals \cite{lacroix2011},  % not the ideal reference here
entangled states of matter \cite{zeng2019quantum},
and, more broadly, fractionalization \cite{Savary2017}.
In \emph{classical} magnetic systems, one of its most striking manifestations is in the formation of non-trivial classical ground-state manifolds \cite{Wannier1950,Henley1987,Henley1989,Moessner1998b,Bramwell2001,Bergman2007}.
In the most extreme case, \emph{local constraints} can generate manifolds that are extensive in size, as in the case of classical spin liquids \cite{Moessner2001,Henley2010,Balents2010}. However, it's also possible to generate \emph{emergent global symmetries}, even emergent continuous symmetries in models that possess only discrete symmetries. This leads to a rich interplay between these emergent symmetries and the true underlying symmetries of the Hamiltonian, an interplay that can have dramatic consequences for the finite-temperature physics of the model. In particular, the accidental nature of the degenerate manifolds that emergent symmetries live in means they are not protected against thermal fluctuations, leaving the door open to thermal \emph{order-by-disorder} \cite{Villain80,Henley1989}.

There have been a number of disparate models that have been shown to exhibit emergent continuous symmetries within their classical ground-state manifold. In two spatial dimensions, a particularly relevant recent example is the Kitaev-Heisenberg model on the honeycomb lattice. Despite the strongly anisotropic nature of the interactions, the model exhibits an emergent continuous $U(1)$ symmetry in its ground-state manifold for all couplings \cite{Price2012,Price2013}. In three spatial dimensions, one of the most well-studied examples is the Hamiltonian describing the rare-earth pyrochlore magnet ErTi$_2$O$_7$ \cite{Champion2004,Savary2012,Zhitomirsky2012,McClarty2014}. Within an extended parameter regime, its ground-state manifold consists of a continuous one-dimensional family of states with the four spins in each tetrahedra pointing within the plane perpendicular to the local $[111]$ axis at each site. The model thus also exhibits an emergent $U(1)$ symmetry, and the corresponding material has been shown to exhibit experimental signatures of quantum order-by-disorder \cite{Ross2014}.

\begin{figure}[b]
    \centering
    \includegraphics[width=\columnwidth]{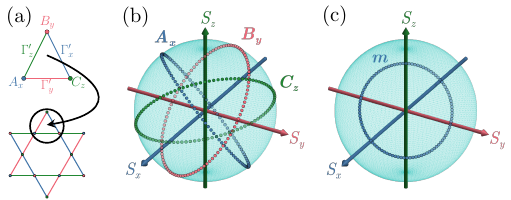}
    \caption{{\bf Emergent symmetries in the $\GP$ model.}
        (a)  Lattice building blocks:
        The kagome lattice as an example lattice geometry built from corner-sharing triangles,
        with three types of bond-directional spin anisotropic interactions.
        (b) Common origin plot:
        Sublattice spins in the ground state manifold of the antiferromagnetic $\GP$-models align along one-dimensional rings in spin space.
        (c) Net magnetization vector in the ground-state manifold lies within the plane perpendicular to the $[111]$ direction.
    }
    \label{fig:antiferromagnetCommonOriginPlots}
\end{figure}

In this manuscript we turn to a family of models, in which an emergent $U(1)$ symmetry arises from an interplay of lattice geometry and bond-directional
interactions -- the classical limit of the $\GP$-model, defined by the bond-directional Hamiltonian
\be
H= \pm \GP \sum_{\avg{i,j}\in \gamma} \sum_{\alpha\neq \gamma} \left( S_i^\alpha S_j^\gamma + S_i^\gamma S_j^\alpha \right) \;
\label{eqn:Ham}
\ee
with the factor of $\pm\GP$ denoting the antiferromagnetic (AFM) and ferromagnetic (FM) versions of the model. This form of symmetric off-diagonal interaction, which possesses only discrete spin and lattice symmetries, is perhaps most well-known as part of the Hamiltonian describing the honeycomb Kitaev materials \cite{Trebst22,Rousochatzakis2024}. In such materials, it is typically small relative to the other symmetry-allowed terms, whereas here we study it in its own right in splendid isolation. Considering the individual properties of such symmetric off-diagonal exchange, our study of the $\GP$-model extends a line of previous work considering classical variants of the Kitaev model \cite{Chandra2010,Sela2014} and the $\Gamma$-model \cite{Rousochatzakis2017,Chern2019},
both of which are known to give rise to classical spin liquids.

The basic physics of the $\GP$-model can be understood by solving its AFM version on a \emph{single} triangle, with one of each bond type, $x$, $y$ and $z$, as given in Fig.~\ref{fig:antiferromagnetCommonOriginPlots}(a). This 3-spin model already gives rise to a ground-state manifold with an emergent continuous U(1) symmetry, as schematically illustrated in Fig.~\ref{fig:antiferromagnetCommonOriginPlots}(b). By combining multiple such triangles together, subject to certain constraints, one can construct lattice geometries in one, two and three spatial dimensions, all of which exhibit the exact same ground-state manifold and associated emergent symmetry. This unique situation allows us to systematically study the interplay between emergent continuous symmetries and underlying discrete symmetries, and the impact of thermal and quantum fluctuations, in one, two and three spatial dimensions. As representative examples, we focus on the two-dimensional kagome and three-dimensional hyperkagome lattices, which both feature an underlying discrete $\mathbb{Z}_6$ symmetry (and only briefly discuss a one-dimensional example). Such a high-order $\mathbb{Z}_N$ symmetry gives rise to non-trivial finite-temperature physics (see Appendix \ref{appendix:otherlattices} for an example of a lattice with an underlying $\mathbb{Z}_2$ symmetry).  

We first discuss, in Sec.~\ref{sec:singletriangle}, the model at the level of a single triangle and the ground states of the ferromagnetic (FM) and antiferromagnetic (AFM) versions of the model. In Sec.~\ref{sec:lattices} we extend the model to full lattices in one, two, and three spatial dimensions by elucidating a set of lattice construction rules that preserve the structure of the AFM ground-state manifold. We then turn to finite temperatures in Sec.~\ref{sec:thermodynamics}. A thermal order-by-disorder calculation indicates that thermal fluctuations
select a discrete subset of maximally non-coplanar ground states. Classical Monte Carlo simulations on the kagome and hyperkagome lattices confirm this thermal order-by-disorder selection, as well as the expected finite-temperature criticality.
In Sec.~\ref{sec:QObD}, we examine the impact of quantum fluctuations and which states they favor.
Finally, Sec.~\ref{sec:discussion} offers some concluding remarks and discussion.

%%%%%%%%%%%%%%%%%%%%%%%%%%%%%%%%%%%%%%%%%%%%
%% Single Triangle
%%%%%%%%%%%%%%%%%%%%%%%%%%%%%%%%%%%%%%%%%%%%

\section{Physics of a Single Triangle}\label{sec:singletriangle}

Let us begin by discussing the physics of the model at the level of a single triangle,
which turns out to provide general guidance to the collective physics of the model.
For a single triangle, there are three sites, which we label as $A_x,B_y,C_z$,
each of which is connected to only two out of the three bond types $\GP_x, \GP_y$ or $\GP_z$
as shown in Fig. \ref{fig:antiferromagnetCommonOriginPlots}(a).
The subscript in the site labels indicates the bond they are not connected to,
e.g.~the $A_x$ site is only connected to $y$ and $z$ bonds.
Writing the bond-directional exchange in matrix form,
\be
\begin{array}{ccc}
    \GP_x=\begin{pmatrix}
              0 & 1 & 1 \\
              1 & 0 & 0 \\
              1 & 0 & 0 \\
          \end{pmatrix}, &
    \GP_y=\begin{pmatrix}
              0 & 1 & 0 \\
              1 & 0 & 1 \\
              0 & 1 & 0 \\
          \end{pmatrix}, &
    \GP_z=\begin{pmatrix}
              0 & 0 & 1 \\
              0 & 0 & 1 \\
              1 & 1 & 0 \\
          \end{pmatrix} .
\end{array}
\ee
one can straightforwardly see the specific symmetric off-diagonal nature of the interactions. Of course, it's important to note that, by a change of local basis, it's also possible to rewrite the above interactions in a different form. For our purposes, the $\GP$-parameterization is the simplest and most transparent form of the Hamiltonian. 

\subsection{Ferromagnetic Model}
\begin{figure}
    \includegraphics[width=\linewidth]{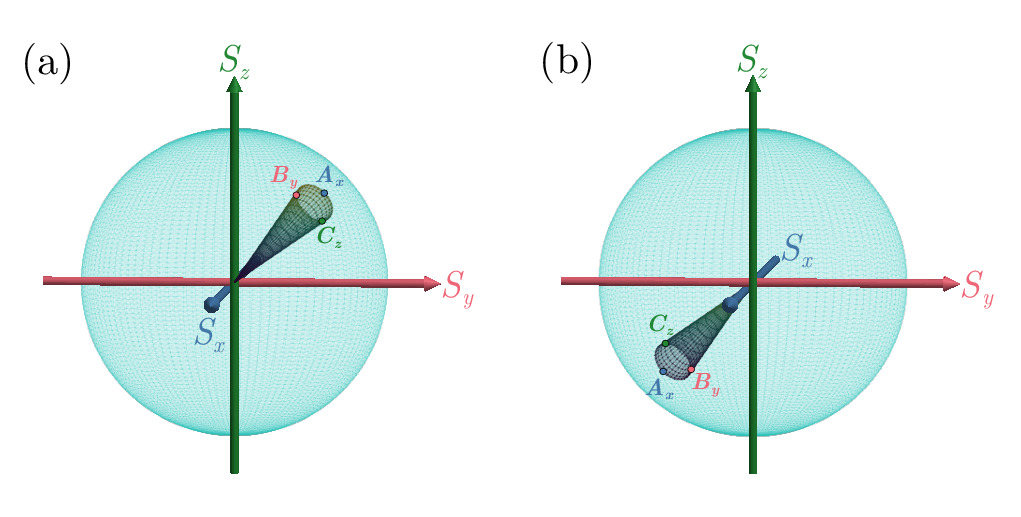}
    \caption{{\bf Ferromagnetic ground state.}
        Common origin plots for the two-fold degenerate ferromagnetic ground state,
        with spins tilted by an angle $\phi=\left(1/2\right)\arctan{\left(\sqrt{2}/5\right)}$ away from
        (a) the $[111]$ direction and (b) the $[\bar{1}\bar{1}\bar{1}]$ direction.
        %Shown is a conceptual rendering.
    }
    \label{fig:originPlotFM}
\end{figure}

When solving this elementary 3-spin model for FM interactions, one finds -- unsurprisingly -- that the classical ground state on a single triangle is two-fold degenerate. Specifically, one finds two possible spin configurations where the three spins are canted either slightly away from the [111] or [$\bar{1}\bar{1}\bar{1}$] direction, as shown in Fig.~\ref{fig:originPlotFM}, by a fixed opening angle $\phi$ toward the $-x,-y,-z$ (or $x,y,z$) axes for sites $A_x,B_y,C_z$.
This allows to explicitly write the FM ground state spin configurations as
\be
\begin{aligned}
    \vec{S}_{A_x} & = \pm \frac{1}{\sqrt{6}} \begin{pmatrix*}[l]
                                                 \sqrt{2}\cos{\phi}-2\sin{\phi}  \\
                                                 \sqrt{2}\cos{\phi} + \sin{\phi} \\
                                                 \sqrt{2}\cos{\phi} + \sin{\phi}
                                             \end{pmatrix*}, \\
    \vec{S}_{B_y} & = \pm \frac{1}{\sqrt{6}} \begin{pmatrix*}[l]
                                                 \sqrt{2}\cos{\phi} + \sin{\phi} \\
                                                 \sqrt{2}\cos{\phi}-2\sin{\phi}  \\
                                                 \sqrt{2}\cos{\phi} + \sin{\phi}
                                             \end{pmatrix*}, \\
    \vec{S}_{C_z} & = \pm \frac{1}{\sqrt{6}} \begin{pmatrix*}[l]
                                                 \sqrt{2}\cos{\phi} + \sin{\phi} \\
                                                 \sqrt{2}\cos{\phi} + \sin{\phi} \\
                                                 \sqrt{2}\cos{\phi}-2\sin{\phi}
                                             \end{pmatrix*} \;,
\end{aligned}
\ee
\noindent
with a fixed angle $\phi=\left(1/2\right)\arctan{\left(\sqrt{2}/5\right)}$.
The corresponding ground-state energy per site is $E_0= - \frac{(1+\sqrt{3})}{2} \GP$.

\subsection{Antiferromagnetic Model}
In contrast, when moving to AFM interactions the 3-spin model shows a notable distinction in exhibiting a degenerate manifold of possible ground-state spin configurations.
This manifold of ground states can be parameterized by a single, continuous angle $\theta$ as
\begin{equation}
    \begin{aligned}
        \vec{S}_{A_x} & = \sqrt{\frac{2}{3}} \begin{pmatrix*}
                                                 -\cos\theta   \\
                                                 +\cos(\theta+{2\pi}/{3}) \\
                                                 +\cos(\theta-{2\pi}/{3})
                                             \end{pmatrix*},  \\
        \vec{S}_{B_y} & = \sqrt{\frac{2}{3}} \begin{pmatrix*}
                                                 +\cos(\theta+{2\pi}/{3})\\
                                                 -\cos(\theta-{2\pi}/{3})  \\
                                                 \cos\theta
                                             \end{pmatrix*}, \\
        \vec{S}_{C_z} & = \sqrt{\frac{2}{3}} \begin{pmatrix*}
                                                 +\cos(\theta-{2\pi}/{3})\\
                                                 \cos\theta \\
                                                 -\cos(\theta+{2\pi}/{3})
                                             \end{pmatrix*} \;,
    \end{aligned}
    \label{eqn:AFMGS}
\end{equation}
with a ground-state energy per site of $E_0=-\GP$ (independent of $\theta$).
The spins $\vec{S}_{A_x},\vec{S}_{B_y},\vec{S}_{C_z}$
lie in planes perpendicular to the [$\bar{1}$11], [1$\bar{1}$1] and [11$\bar{1}$]
axes respectively, as illustrated in Fig.~\ref{fig:antiferromagnetCommonOriginPlots}(b).
Local site-dependent rotations about these three axes naturally shift the angle $\theta$, transforming one ground state into another --
generating a continuous $U(1)$ symmetry within the ground-state manifold.

Importantly, for any choice of the angle $\theta$ these states exhibit a finite net magnetization, with the magnetization vector, $\vec{m}=\frac{1}{3}\sum_i \vec{S}_i$, lying in the plane perpendicular to the $[111]$ direction and given by
\be
\vec{m}= -\frac{2}{3} \sqrt{\frac{2}{3}} \begin{pmatrix} \cos\theta \\ \cos\left(\theta - \frac{2\pi}{3} \right) \\ \cos\left(\theta + \frac{2\pi}{3} \right) \end{pmatrix} ,
\label{eqn:AFMMagnetizationVector}
\ee
with a fixed magnitude $\abs{\vec{m}}=2/3$. 
Though all ground states possess the same fixed net magnetization, they turn out to differ in their scalar spin chirality
\be
\chi=\vec{S}_{A_x} \cdot \left( \vec{S}_{B_y} \times \vec{S}_{C_z} \right)=\frac{2}{3} \sqrt{\frac{2}{3}}\cos \left(3\theta\right) \,.
\label{eqn:chirality}
\ee
This chirality exhibits a $\cos(3\theta)$ periodicity and for almost all angles describe a non-coplanar state
(but $\theta=m\pi/6$ with $m=1,3,5,7,9,11$), see also Fig.~\ref{fig:thermalOrderByDisorder}(b) below.

\subsection{Emergent U(1) symmetry}

\begin{figure}
    \centering
    \includegraphics{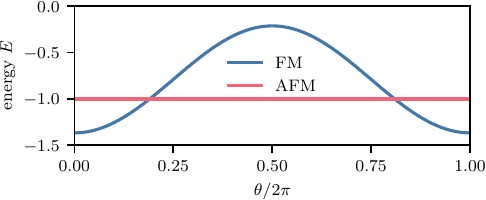}
    \caption{{\bf Absence of a continuous symmetry} in the Hamiltonian
        demonstrated by comparing the energies of representative ground states of the AFM and FM models
        as a function of the $U(1)$ rotation that traverses the AFM ground-state manifold.}
    \label{fig:energyFMvsAFM}
\end{figure}

The continuous $U(1)$ symmetry present in the ground-state manifold of the AFM model might be rather surprising given that the Hamiltonian, as written in Eq.~\eqref{eqn:Ham}, does not obviously possess any continuous symmetries. However, it could be that, given a suitably judicious choice of basis transformation, the Hamiltonian could, in fact, be rewritten in an explicitly $U(1)$ symmetric form. To rule out such an underlying, hidden symmetry one can simply pick a representative spin configuration from the set of FM and AFM ground states and track their energies as one performs the site-dependent rotations that move you within the AFM ground-state manifold. As shown in Figure \ref{fig:energyFMvsAFM}, the energy of the AFM ground state, as expected, stays fixed, while the energy of the FM ground state continuously changes.

We can thus safely conclude that the $U(1)$ symmetry within the AFM ground-state manifold is \emph{emergent}, a result of the accidental degeneracy, and not related to any underlying symmetry of the Hamiltonian. In other words, the Hamiltonian itself possesses only discrete spin and lattice symmetries, and the $U(1)$ symmetry emerges only in the zero-temperature ground-state manifold.

\FloatBarrier

%%%%%%%%%%%%%%%%%%%%%%%%%%%%%%%%%%%%%%%%%%%%
%% "Lego" rules
%%%%%%%%%%%%%%%%%%%%%%%%%%%%%%%%%%%%%%%%%%%%

\section{Lattice Construction Rules}\label{sec:lattices}

Having explored the physics of the model on a single triangle, we now turn to combining such triangles together to construct full lattice geometries in one, two, and three spatial dimensions. 
 
The task of combining two triangles together can be done by either sharing a site i.e. corner-sharing or a bond i.e. edge-sharing.
The key finding here is that, by following a specific set of rules, ensuring that shared site(s) are of the same type, it is possible to preserve the exact same ground-state manifold and associated emergent $U(1)$ symmetry as in the single triangle. Thus, one can \emph{systematically} build lattice models from scratch that will exhibit an emergent continuous symmetry and study the consequences of the same emergent symmetry in different spatial dimensionalities.

\subsection{Corner-sharing}
For lattices of corner-sharing triangles, in which shared sites are of the same type, the Hamiltonian can be written as a sum over individual triangles
\be
H=\sum_{\triangle} \mathcal{H}_\triangle \,.
\ee
For classical spins, all terms appearing in the Hamiltonian commute and so the ground state of the full Hamiltonian is simply the ground state of each individual triangle \cite{HanYan17}.

As an example, consider the simplest case in which a single site is shared between just two triangles. Labelling the sites as $A_x, B_y, C_z$ and $A_x\pr, B_y\pr, C_z\pr$,
the resulting Hamiltonian is simply the sum $H+H\pr$.
The ground state for each triangle can be defined by a single parameter $\theta$ and $\theta\pr$
which minimizes the individual Hamiltonian for each respective triangle.
Assuming, without loss of generality, that it is $A_x$ that is shared between the triangles sets $A_x=A_x\pr$. Following the definition of $\vec{S}_{A_x}$ in Eq.~\ref{eqn:AFMGS}, this necessarily implies that $\theta=\theta\pr$ (modulo $2\pi$), and the two triangles together thus possess a single continuous emergent $U(1)$ symmetry.

\begin{figure}
    \centering
    \includegraphics[width=\columnwidth]{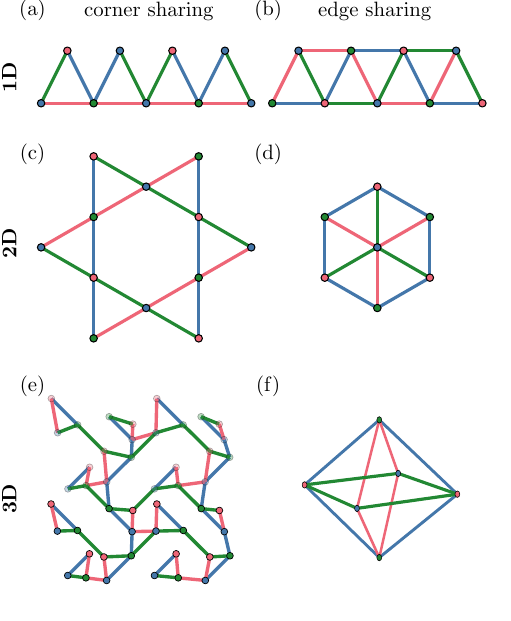}
    \caption{{\bf Lattice construction.}
        Shown are a number of one-, two- and three-dimensional lattice geometries
        that allow to us to define an AFM $\GP$-model with an emergent U(1) symmetry.
        The left column shows corner-sharing lattice geometries including
        (a) a corner-sharing triangular chain, (c) the kagome and (e) the hyperkagome lattices.
        The right column shows edge-sharing lattice geometries including
        (b) an edge-sharing trianglular chain, (d) the triangular and (f) the octahedral lattices.
    }
    \label{fig:latticeConstruction}
\end{figure}

Example lattice geometries that can be constructed using this method of
connecting corner-sharing triangles, such that sites of the same type are shared, are illustrated in 
the left column Figure \ref{fig:latticeConstruction},
including the much studied kagome and hyperkagome lattices, which will also serve
as principal examples in our further analysis.

\subsection{Edge-sharing}

The edge-sharing scenario refers to the case where two sites of the same type (along an edge)
are shared between two triangles that are connected within a larger lattice geometry. In other words, bonds of the same type must be shared between triangles. A few examples of lattices that can be constructed using this edge-sharing approach are visualized in the right column of Figure \ref{fig:latticeConstruction}, which includes the triangular lattice with an enlarged unit cell
as a principal example. We have confirmed that all of these examples exhibit the same ground state manifold and emergent $U(1)$ symmetry for the AFM $\GP$-model. However, more generally, for edge-sharing lattices the Hamiltonian does not have a neat representation as a sum over independent triangles, making it difficult to definitively rule out degenerate magnetic orders with larger unit cells on other edge-sharing lattices.  

\subsection{Bottom-up approach}
While we have not carried out a complete classification of all possible lattice geometries with a built-in $U(1)$ symmetry for their respective AFM $\GP$-model, we conjecture that the key properties required are that
(i) every site must be connected by only two out of the three bond types,
and (ii) the graph must be assembled from connecting triangles.
These properties can be fulfilled by all graphs that satisfy two conditions:
(a) the graph must be three-colorable, and
(b) the graph must be decomposable into triangles with repeating edges.
The reasoning behind these conditions is as follows,
since every site on the lattice must be missing one type of bond
and there are only three types of bonds, the lattice must be
three colorable and since the graphs must be assembled from
triangles using the construction methods mentioned above, it must remain
triangle decomposable when accounting for repeating edges as
in the case for edge-sharing lattices.

%%%%%%%%%%%%%%%%%%%%%%%%%%%%%%%%%%%%%%%%%%%%
%% Thermodynamics
%%%%%%%%%%%%%%%%%%%%%%%%%%%%%%%%%%%%%%%%%%%%

\section{Thermodynamics}
\label{sec:thermodynamics}

With a family of lattice geometries in one, two, and three spatial dimensions, for which we know
-- by construction -- that the ground-state manifold of the AFM $\GP$-model exhibits a continuous
$U(1)$ symmetry, we now move to their finite-temperature physics. In particular, we investigate
the impact of thermal fluctuations on this emergent symmetry and its dependence on spatial
dimensionality. While we argue that our results will hold on general grounds (supported by
arguments from renormalization group calculations), we will also study two models in great detail
-- the  $\GP$-model on the kagome and hyperkagome lattices as two particularly rich representative examples in 2D and 3D.

%%%%%%%%%%%%%%%%%%%%%%%%%%%%%%%%%%%%%%%%%%%%
\subsection{Thermal Order-by-Disorder} \label{sec:TObD}
%%%%%%%%%%%%%%%%%%%%%%%%%%%%%%%%%%%%%%%%%%%%

The emergent $U(1)$ symmetry is a direct manifestation of the formation of a degenerate, one-dimensional ground-state
manifold, i.e.~of equal energy spin configurations at zero temperature.
However, since the symmetry is emergent, rather than intrinsic to the Hamiltonian, it is not capable
of protecting this ground-state manifold against fluctuations.
This lack of protection is reflected in the common description of the ground-state degeneracy as ``accidental".
In fact, even the simplest of perturbations -- thermal fluctuations -- are potent enough to lift this degeneracy, by introducing small variations in the free energy of different configurations.
This has important consequences for the thermodynamics of the model. At small but finite temperatures, thermal fluctuations will select the configurations which minimize the free energy, i.e.~those with the largest entropy. Such an entropy-driven selection mechanism is referred to as \emph{thermal order-by-disorder} \cite{Villain80}.

\begin{figure*}
    \centering
    \includegraphics{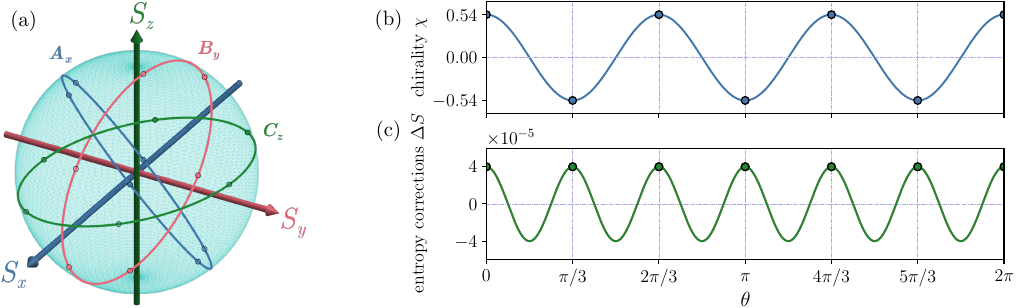}
    \caption{{\bf $\mathbb{Z}_6$ ordering induced by thermal order-by-disorder in the antiferromagnetic ground-state manifold.}
        Figure (a) schematically illustrates the thermal order-by-disorder selection of six states (solid points)
        amongst the U(1) symmetric ground-state manifold (rings).
        Panel (b) shows the chirality (eq.~\eqref{eqn:chirality}) of the ground states and (c) the entropy corrections (eq.~\eqref{eqn:EntropyCorrections}) for the kagome and hyperkagome lattices.
        States with a maxima in $\Delta S$ correspond to free energy minima, and end up being selected due to thermal fluctuations.
        These states are maximally chiral states i.e. they have the maximum magnitude of $\chi = 2/3\sqrt{2/3} \approx 0.544$. % 331053951817$.
    }
    \label{fig:thermalOrderByDisorder}
\end{figure*}

The entropy correction  due to thermal fluctuations can be explicitly computed by expanding in fluctuations about the manifold of ground-state spin configurations. We begin by rotating the sublattice spins into a local basis aligned along the local $z$-axis, $\tilde{\vec{S}}_i=R_i \vec{S}_i$.
We can then expand in two orthogonal fluctuations $\delta u_{i}, \delta v_{i}$ about this ordered direction
\begin{align}
    \notag    \tilde{\vec{S}}_{i} & =\begin{pmatrix}
                                         \delta u_{i}                                 \\
                                         \delta v_{i}                                 \\
                                         \sqrt{1 -\delta u^{2}_{i} -\delta v^{2}_{i}} \\
                                     \end{pmatrix} \\
                                  & \approx
    \begin{pmatrix}
        \delta u_{i}                                                 \\
        \delta v_{i}                                                 \\
        1-\frac{1}{2}\left(\delta u^{2}_{i} +\delta v^{2}_{i}\right) \\
    \end{pmatrix} \,.
\end{align}
\noindent
Substituting this expansion of the spin into the Hamiltonian yields
\be
H=E_0 + H^{(2)} + \dots \,,
\ee
\noindent
where $E_0$ is the classical ground-state energy, all linear fluctuation terms vanish since the selected state is a ground
state of the system, and $H^{(2)}$ contains all of the terms quadratic in the fluctuations.

The problem now involves the diagonalization of $H^{(2)}$, which can be straightforwardly done in Fourier space by taking a
Fourier transform of the fluctuations and collecting them in
a fluctuation vector $\tilde{u}(q)=\left[\delta u(q), \delta v(q)\right]^T $. This allows us to write $H^{(2)}$ as
\be
H^{(2)}=\frac{1}{2}\sum_q \tilde{u}(-q)^T \cdot M(q) \cdot \tilde{u}(q) \,,
\ee
\noindent
where $M(q)$ is built from $2N_s \times 2N_s$ blocks, with $N_s$ the number of sublattice spins.
We can diagonalize the matrix $M(q)$ using an orthogonal transformation $U=(U^T)^{-1}$, resulting in
\be
H^{(2)}=\frac{1}{2} \sum_q \sum_{n=1}^{2N_s} \kappa_{n,q}\nu_{n,q}\nu_{n,-q}
\ee
\noindent
where $\nu_{n,q}$ represent the normal modes of the system, and $\kappa_{n,q}$ are the respective eigenvalues.

Now that we have the fluctuation spectrum up to quadratic order, we can approximate the classical partition function as
\begin{align}
    \begin{split}
        \notag       \mathcal{Z}^{(2)} ={} & \left( \frac{1}{\sqrt{2\pi}}\right)^{2N_s} \exp \left( \frac{-E_0}{T}\right) \int{\prod_{n=1}^{2N_s} \prod_q d \nu_{n,q}} \\
                                           & \exp \left( -\frac{1}{2T} \sum_{n=1}^{2N_s} \sum_q \kappa_{n,q} \nu_{n,q} \nu_{n,-q}\right)
    \end{split} \\
    ={} & \exp\left({\frac{-E_0}{T}} \right) \prod_{n=1}^{2N_s} \prod_q
    \left( \sqrt{\frac{T}{\kappa_{n,q}}}\right) \;.
\end{align}

In the low temperature limit where $T \rightarrow 0$ the free energy
can then be approximated by
\be
\mathcal{F}=E_0 + \frac{T}{2} \sum_{nq} \ln \kappa_{n,q} - NT\ln T \,,
\ee
and the entropy per spin at low temperatures becomes
\begin{align}
    \notag   \frac{S}{N} & = -\frac{1}{N}\frac{\partial F}{\partial T}              \\
                         & = \ln T + 1 -\frac{1}{2N} \sum_{nq} \ln \kappa_{n,q} \;, %\ln( \det({M(q)})) 
\end{align}
\noindent
where $N$ is the number of sites.
Finally, we are interested not in the absolute value of the entropy, but rather in the differences in entropy between different states within the ground-state manifold. We therefore subtract off the average entropy within the manifold and focus on the entropy splitting $\Delta S(\theta) = S(\theta) - \bar{S}(\theta)$.

For our example lattice geometries, the kagome and hyperkagome lattices,
we find that this correction takes the form
\be
\Delta S(\theta) = \delta_{\rm thermal} \cdot \cos \left(6\theta\right) \;
\label{eqn:EntropyCorrections}
\ee
with a coefficient $\delta_{\rm thermal}= 4\times 10^{-5}$ for both lattices.
Notably, this points to a six-fold periodic modulation of the entropy correction as plotted in Fig.~\ref{fig:thermalOrderByDisorder}(c)
and a $\mathbb{Z}_6$ symmetry describing the resulting six minima of the free energy at angles
\[
    \theta = m\frac{\pi}{3} \quad\quad {\rm with} \quad\quad m=0,1,2,3,4,5 \,.
\]
This implies that, at finite temperatures, the continuous $U(1)$ symmetry is broken down to a discrete $\mathbb{Z}_6$ symmetry
and six specific spin configurations (corresponding to the angles $\theta$ above) are selected in a thermal order-by-disorder process.
The six states are indicated by the six solid points in Fig.~~\ref{fig:thermalOrderByDisorder}(a).
Plugging the six relevant angles into the expression of the spin chirality of Eq.~\eqref{eqn:chirality}, we note that
the maxima in entropy, i.e.~the minima of the free energy, correspond to the six states
with the maximum magnitude of $\chi$, as shown in Fig.~\ref{fig:thermalOrderByDisorder}.
Thus, thermal fluctuations select the states that are maximally non-coplanar.

%%%%%%%%%%%%%%%%%%%%%%%%%%%%%%%%%%%%%%%%%%%%
%% Finite Temperature
%%%%%%%%%%%%%%%%%%%%%%%%%%%%%%%%%%%%%%%%%%%%

%%%%%%%%%%%%%%%%%%%%%%%%%%%%%%%%%%%%%%%%%%%%
\subsection{Finite-Temperature Criticality}
%%%%%%%%%%%%%%%%%%%%%%%%%%%%%%%%%%%%%%%%%%%%

We have shown in the previous subsection that thermal fluctuations favor a discrete subset of the AFM ground states via the thermal order-by-disorder mechanism. This six-fold selection of states reflects the fact that the underlying symmetry of the Hamiltonian at play here, for the kagome and hyperkagome lattices, is in fact a discrete $\mathbb{Z}_6$ symmetry, a simple consequence of the fact that a combined $2\pi/3$ spin rotation about the $[111]$ axis, $C_3$ lattice rotation and spin inversion ($\vec{S}_i\rightarrow -\vec{S}_i$) leaves the Hamiltonian unchanged, rather than the accidental $U(1)$ symmetry observed within the ground-state manifold. The question thus arises of how these symmetries are broken at finite temperatures. Fortunately, the outcome is well-known from renormalization group calculations for the XY universality class with $\mathbb{Z}_6$ clock anisotropy \cite{Jose77,Masaki00}, which we summarize in the following (though such calculations were carried out for planar $O(2)$ spins, one expects the same universal physics to occur for $O(3)$ spins).

%%%%%%%%%%%%%%%%%%%%%%%%%%%%%%%%%%%%%%%%%%%%
\subsubsection*{Two spatial dimensions}
%%%%%%%%%%%%%%%%%%%%%%%%%%%%%%%%%%%%%%%%%%%%

In two spatial dimensions, such as the kagome lattice case, there are two distinct finite temperature phase transitions \cite{Jose77}. At high temperatures, the system is of course in a paramagnetic state. At low temperatures, the clock anisotropy is relevant and drives the system into a long-range ordered ground state breaking the discrete $\mathbb{Z}_6$ symmetry. However, at intermediate temperatures, the clock anisotropy becomes irrelevant and there is an intermediate critical XY phase with quasi-long range order and power-law decaying correlations. This intermediate phase is flanked by two critical temperatures, an upper $T_c^h$ separating it from the high-temperature paramagnet and a lower $T_c^l$ separating it from the low-temperature long-range ordered phase. Both transitions are within the Berezinskii-Kosterlitz-Thouless (BKT) universality class, with the expected sequence of transitions illustrated in Fig.~$\ref{fig:transition2D3D}$(a).

%%%%%%%%%%%%%%%%%%%%%%%%%%%%%%%%%%%%%%%%%%%%
\subsubsection*{Three spatial dimensions}
%%%%%%%%%%%%%%%%%%%%%%%%%%%%%%%%%%%%%%%%%%%%

\begin{figure}[t]
    \centering
    \includegraphics[width=\linewidth]{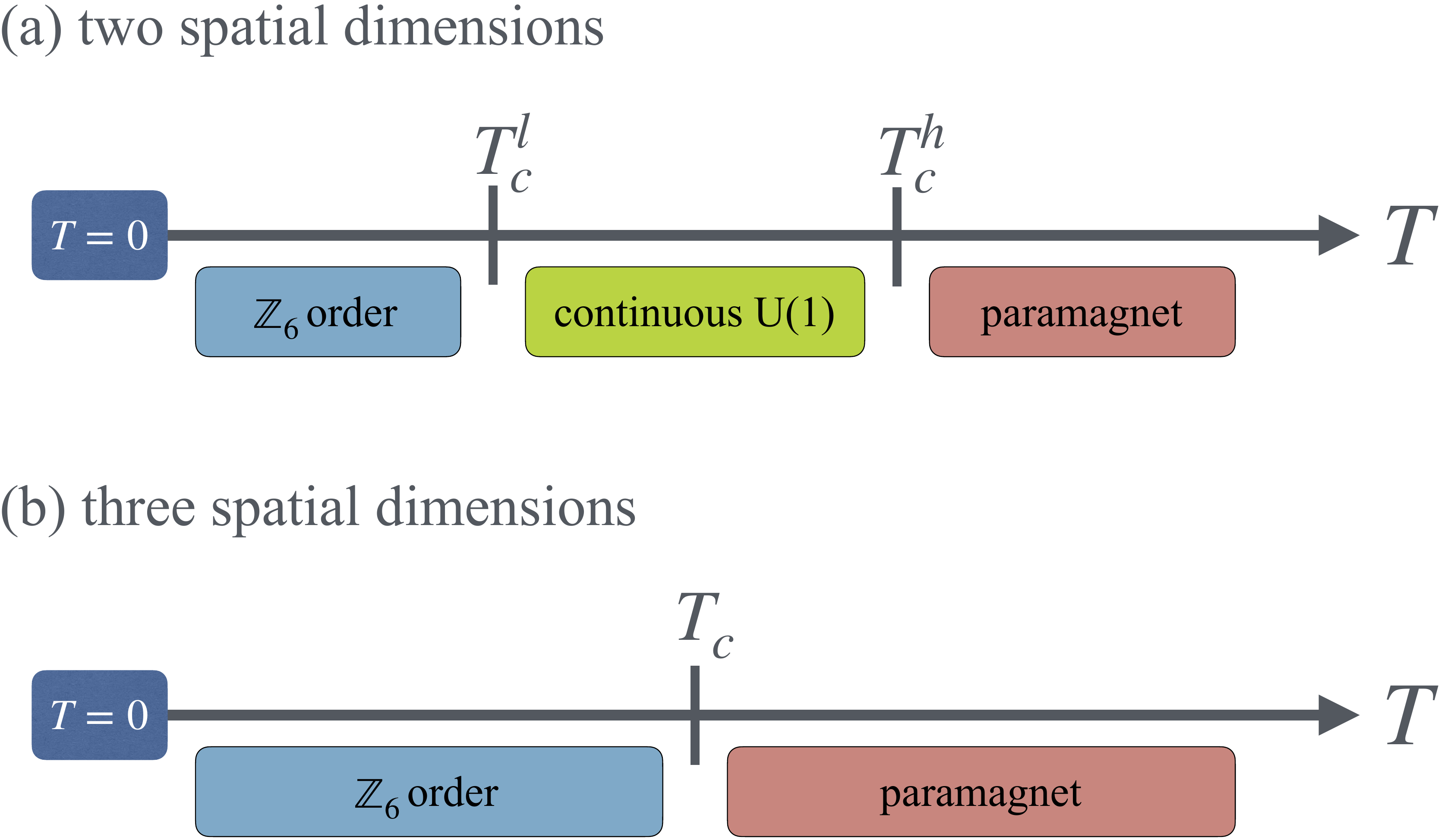}
    \caption{{\bf Schematic of finite-temperature criticality for two and three-dimensional systems}.
    Illustrated are the schematic phase diagrams for the anti-ferromagnetic $\GP$ model
    on (a) two-dimensional lattices such as kagome and (b) three-dimensional lattices
    such as hyperkagome.
    In the two-dimensional case, there is an intermediate continuous U(1)
    phase (XY phase) that exists at finite temperatures even after properly accounting
    for thermal fluctuations, sandwiched between two BKT transitions at $T_c^h$ and $T_c^l$. 
    However, in the 3-dimensional case, the U(1) phase gets destroyed by
    thermal fluctuations at all temperatures in the thermodynamic limit,
    leading to the absence of any intermediate phase.
    }
    \label{fig:transition2D3D}
\end{figure}

In three dimensions, such as the hyperkagome lattice case, the historical development of the phase diagram was more complicated. It is now believed that there is just a single direct transition from the high-temperature paramagnetic phase to the long-ranged ordered $\mathbb{Z}_6$ symmetry-breaking phase \cite{Masaki00}. The transition belongs to the three-dimensional XY universality class. This scenario is illustrated in Fig.~\ref{fig:transition2D3D}(b). However, for temperatures slightly below the transition, the relevance of the clock anisotropy is small and thus, in any finite-sized system, the ordered phase here actually resembles a symmetric XY phase. This implies that in finite-sized systems, within the long-range ordered phase, there will be a crossover from U(1) symmetric correlations to true $\mathbb{Z}_6$ symmetry-breaking behavior at lower temperatures (with the window of $U(1)$ symmetric correlations shrinking with increasing system size).

There is thus a stark contrast between the two- and three-dimensional cases, the former possesses two finite-temperature phase transitions with an intermediate XY phase while the latter possesses just a single finite-temperature transition and thermal crossover.

%%%%%%%%%%%%%%%%%%%%%%%%%%%%%%%%%%%%%%%%%%%%
%% Numerical simulations
%%%%%%%%%%%%%%%%%%%%%%%%%%%%%%%%%%%%%%%%%%%%

%%%%%%%%%%%%%%%%%%%%%%%%%%%%%%%%%%%%%%%%%%%%
\subsection{Classical Monte-Carlo Simulations}
%%%%%%%%%%%%%%%%%%%%%%%%%%%%%%%%%%%%%%%%%%%%

Now that we know what to expect at finite temperatures, we turn to classical Markov-Chain Monte Carlo techniques \cite{LandauBinder} to investigate the thermodynamics of the AFM $\Gamma^\prime$-model and confirm these expectations. The technical details of the Monte Carlo simulations are discussed in Appendix \ref{appendix:MonteCarlo}.

Before proceeding to specific examples, we first define the useful quantities of interest. Defining the total magnetization vector as $\vec{m}=\sum_i \vec{S}_i/N$, with $N$ the number of sites, we can then define the magnetization $m=\abs{\vec{m}}$. By projecting the total magnetization vector into the plane perpendicular to the $[111]$ direction we obtain the planar magnetization vector $\vec{m}^p=(m^p_{\tilde{x}},m^p_{\tilde{y}})^T$, with $\tilde{x}$ and $\tilde{y}$ forming a two-dimensional coordinate system in the place perpendicular to $[111]$. Within the AFM ground-state manifold $m=m^p=2/3$.

We can define an order parameter for the $\mathbb{Z}_6$ symmetry broken phase as
\be
z_6=\cos(6\theta) \,,
\label{eqn:z6}
\ee
where $\theta$ is the angle of the planar magnetization vector within the $\tilde{x}$-$\tilde{y}$ plane (defined such that $\theta=0$ corresponds to a maximally non-coplanar ground state with $\chi=+2/3\sqrt{2/3} \approx 0.544$). Within the XY phase, $\theta$ is uniformly distributed and thus the average order parameter vanishes, $\bar{z}_6=0$,
while in the $\mathbb{Z}_6$ symmetry broken phase we expect from the thermal order-by-disorder calculations that $\bar{z}_6\goto 1$.

\begin{figure}
    \centering
    \includegraphics[width=\linewidth]{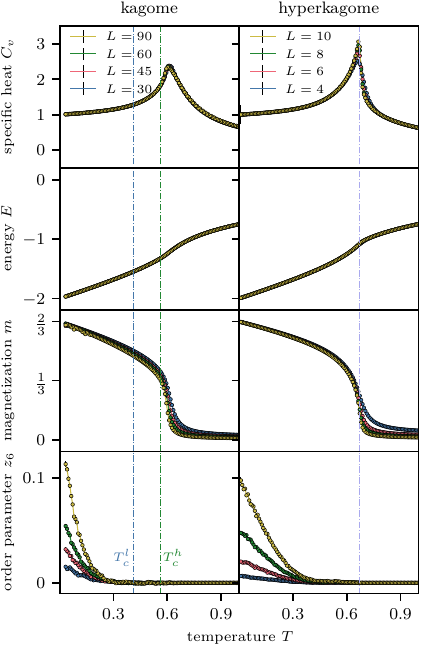}
    \caption{{\bf Thermodynamics.}
        Shown are numerical results from classical Monte Carlo simulations
        for the kagome (left) and hyperkagome (right) lattices for system
        sizes $L=30,45,60,90$ and $L=$ 4,6,8,10 respectively,
        i.e.\ up to a total of $N=$ 24,300 and $N=$ 12,000 spins, respectively.
        The top panel shows the specific heat, the second panel
        shows the energy, the third panel shows the
        magnetization and the bottom panel
        shows the $z_6$ order parameter of Eq.~\eqref{eqn:z6}).
        While the peak in the specific heat points to a transition out
        of the high-temperature paramagnet the latter indicates a transition
        to a $\mathbb{Z}_6$ ordered phase.
        The dash-dotted lines indicate the transition temperatures $T_c^l$
        and $T_c^h$ extracted from the scaling behavior of the planar
        magnetization in Fig.~\ref{fig:etaVsT} below.
        The data is obtained as averages over $10^9$ measurements.
    }
    \label{fig:MonteCarlo}
\end{figure}

%%%%%%%%%%%%%%%%%%%%%%%%%%%%%%%%%%%%%%%%%%%%
\subsubsection*{Two dimensions: Kagome lattice}
%%%%%%%%%%%%%%%%%%%%%%%%%%%%%%%%%%%%%%%%%%%%

An overview of the results for the two-dimesional kagome lattice case is shown in the left panels of Fig.~\ref{fig:MonteCarlo}. The specific heat exhibits a single peak that does not appear to scale with system size, at which a finite magnetization begins to onset. Such behavior is consistent with a BKT transition from the high-temperature paramagnet to an intermediate XY phase. The existence of a second transition is not visible in the specific heat as the U(1) and $\mathbb{Z}_6$ states have the same energy and only differ in their free energy.
It is instead indicated by the development of a small but finite $z_6$ order parameter at low temperatures.

\begin{figure}
    \includegraphics[width=\columnwidth]{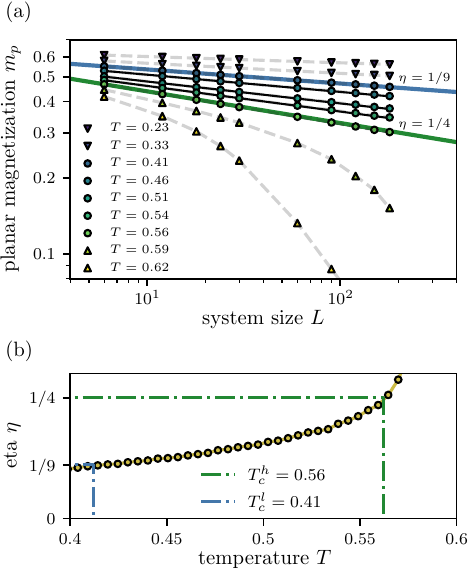}
    \caption{{\bf Scaling behavior of planar magnetization.}
        (a) Curves of the planar magnetization $m^p(L)$ versus linear system size $L$ corresponding
        to select temperatures on a log-log scale for a kagome system.
        The slope of a linear fit defines the exponent $\eta$ in Eq.~\eqref{eqn:mpscaling},
        with the curves between $\eta=1/4$ and $\eta=1/9$ defining the two transition temperatures $T_c^h$ and $T_c^l$
        bounding the extent of the intermediate $U(1)$ phase.
        (b) The exponent $\eta$ as a function of temperature with the intersection of the curve
        with $\eta=1/9,1/4$ used to pinpoint the two transition temperatures $T_c^h$ and $T_c^l$.
    }
    \label{fig:etaVsT}
\end{figure}

To verify the existence of an intermediate XY phase, and to obtain its upper and lower critical temperatures, we turn to an analysis of the finite-size scaling behavior of the magnetization. Within the XY phase, the planar magnetization $m^p(L)$ for a system of length $L$ scales as
\be
m^p(L) \propto L^{-\frac{\eta}{2}} \label{eqn:mpscaling}
\ee
where $\eta$ is the critical exponent \cite{Challa1986}. For the upper transition, it is expected that $\eta=1/4$, while for the lower transition, $\eta=1/9$ \cite{Jose77}. At each temperature, one can obtain $\eta$ as the slope of a linear fit of $\log(m^p(L))$ versus $\log(L)$.
This is indicated in Fig.~\ref{fig:etaVsT}(a), which shows a subset of these curves for a range of temperatures, as well as the extracted $\eta$ as a function of temperature in its panel (b). We clearly observe an intermediate temperature region with power-law scaling of the planar magnetization and can estimate the critical temperatures as $T_c^h \approx 0.56$ and $T_c^l \approx0.41$ (with a precise error estimate hard to determine). 

\begin{figure}
    \includegraphics[width=\columnwidth]{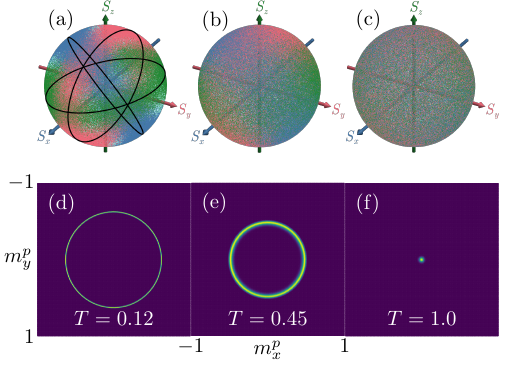}
    \caption{{\bf Visualization of the intermediate U(1) phase.}
        (a)-(c) Common origin plots
        at various temperatures and Figures (d)-(f) show
        the corresponding phase space diagram for the planar magnetization
        vector which is computed as a histogram of the samples taken during
        a Monte Carlo simulation for a kagome system with $L=30$.
        On the left, we see the low temperature ($T=0.12$)
        phase with $m_p \approx 2/3$ as the radius with corresponding
        rings in the common origin plot. In the center figures,
        we see the U(1) phase ($T=0.45$) and on the right we see the
        high temperature ($T=1.0$) paramagnetic phase with $m_p \approx 0$.
        Each common origin plot is a combination of spin configuration snapshots
        and each histogram contains $10^8$ samples. They are both computed from
        the same set of 100 independent runs.
    }
    \label{fig:planarMagnetizationAndCOP}
\end{figure}

The existence of an emergent $U(1)$ symmetry for the intermediate phase can be visualized in a straight-forward manner by creating a histogram of $\vec{m}^p$ at a fixed temperature, as shown in Fig.~\ref{fig:planarMagnetizationAndCOP}. The planar magnetization vector lies on a clear symmetric ``ring", with a radius $m^p$ that continuously increases with decreasing temperature below $T_c^h$.

%%%%%%%%%%%%%%%%%%%%%%%%%%%%%%%%%%%%%%%%%%%%

Note that the extracted upper critical temperature is just below the peak in the specific heat, as expected for a BKT transition. The lower critical temperature is above the regime in our simulations in which we observe a finite $z_6$. Furthermore, the magnitude of $z_6$, even at the lowest temperatures, is just over $10\%$ of its expected saturated value. This is also evident in the histogram of $\vec{m}^p$ at $T=0.12$, well below the lower critical temperature, shown in Fig.~\ref{fig:planarMagnetizationAndCOP}(d), which still maintains approximate circular symmetry. We ascribe this extremely slow and weak onset of the $\mathbb{Z}_6$ order parameter to the extremely small entropy differences between the different states within the AFM ground-state manifold, being on the order of $10^{-5}$ per site (see Fig.~\ref{fig:thermalOrderByDisorder}). In fact, the simulations required an unusually large number of Monte Carlo measurements, about 1 billion ($10^9$) sweeps, before the relatively clean $z_6$ data in Fig.~\ref{fig:MonteCarlo} could be obtained in the first place.

\begin{figure}
    \centering
    \includegraphics[width=\linewidth]{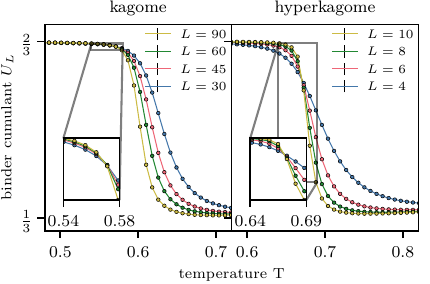}
    \caption{{\bf Binder cumulant at U(1) transition.}
        Shown is the Binder cumulant $U_L$ (equation \ref{eqn:binder})
        and its crossing point from Monte Carlo simulations of
        kagome and hyperkagome lattices. The crossing point indiciates
        the U(1) transition temperature.
        For each temperature point $10^9$ samples are averaged over,
        resulting in statistical error bars that are smaller than the symbol sizes.
    }
    \label{fig:BinderCumulant}
\end{figure}

An alternative method for pinpointing the upper critical temperature is to use the crossing of the Binder cumulant defined as
\be
U_L=1-\frac{\langle m^4\rangle_L}{3 \langle m^2\rangle_L^2} \,.
\label{eqn:binder}
\ee
It is well known that the crossing point indicates a transition from a phase with zero magnetization to one with finite magnetization \cite{Binder1981}. Fig.~\ref{fig:BinderCumulant}(a) shows the crossing point for the kagome lattice with an extracted transition temperature of $T_c^h=0.560(3)$, in good agreement with the value extracted from the finite-sized scaling of $m_L^p$ with $\eta$ (note that a finite-size drift is taken into account by extrapolating
to the thermodynamic limit using
the method described in Ref. \cite{Dolfi2014}).

%%%%%%%%%%%%%%%%%%%%%%%%%%%%%%%%%%%%%%%%%%%%
\subsubsection*{Three dimensions: Hyperkagome lattice}
%%%%%%%%%%%%%%%%%%%%%%%%%%%%%%%%%%%%%%%%%%%%

An overview of the results for the three-dimensional hyperkagome lattice is shown in the right panels of Fig.~\ref{fig:MonteCarlo}. In this case, there is a system-size dependent peak in the specific heat whose peak height systemically increases with increasing system size. At the same temperature, there is an onset of finite magnetization. This temperature marks the transition from the high-temperature paramagnetic phase into the low-temperature $\mathbb{Z}_6$ symmetry broken phase. Its precise location can again be pinpointed by examining the crossing point of the Binder cumulant for the magnetization, as defined in Eq.~\eqref{eqn:binder}. As shown in Fig.~\ref{fig:BinderCumulant}(b), the crossing occurs at $T_c=0.6692(2)$.

As expected, the $z_6$ order parameter does not onset at the same temperature as the magnetization. Below $T_c$, there is a crossover from symmetric U(1) correlations, with finite magnetization and negligible $z_6$, to true $\mathbb{Z}_6$ symmetry breaking behavior, with both finite magnetization and finite $z_6$ order parameter.

\subsection{One-dimensional scenario}

\begin{figure}
    \centering
    \includegraphics[width=\linewidth]{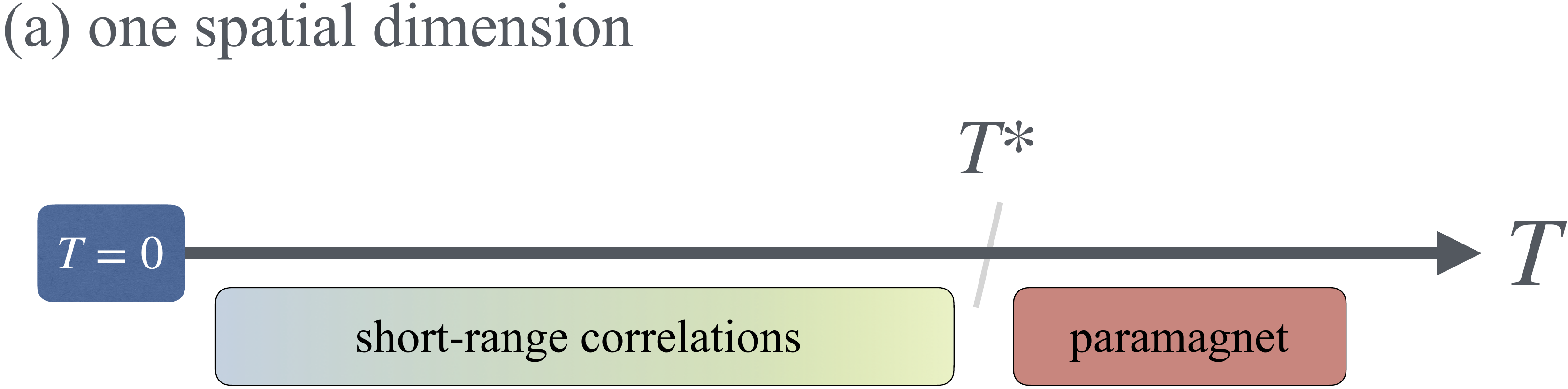}
    \vskip 5mm
    \includegraphics[width=\linewidth]{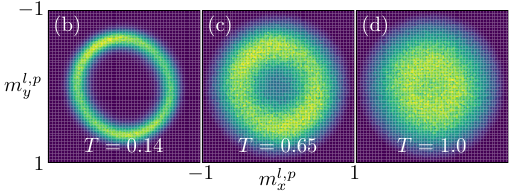}
    \vskip 5mm
    \includegraphics[width=\linewidth]{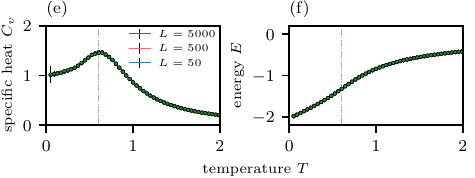}
    \caption{{\bf Finite temperature physics for one-dimensional systems.}
        (a) Schematic phase diagram. In the one-dimensional case, there is no finite-temperature phase transition. There is only a thermal crossover at $T^\star$, from a fully paramagnetic regime to one with short-range spin-spin correlations. 
        (b)-(d) Histograms of the local planar magnetization, i.e.~the planar magnetization of a single triangle. 
        (e)-(f) Specific heat and energy as a function of temperature. The broad bump in the heat capacity indicates the location of the expected thermal crossover. The Monte Carlo simulations are for system sizes $L =  50, 500, 5000$,
        i.e.\ up to a total of $N = 30,000$ spins. 
        The data is obtained as averages over $10^6$ measurements. 
        }
    \label{fig:1DOverview}
\end{figure}

To round off our discussion of the thermodynamic behavior on spatial dimensionality, we provide a brief overview of results for the one-dimensional edge-sharing chain of Fig.~\ref{fig:latticeConstruction}(b). Our analytical expectation, based on the 1D nature of the model, is that no true long-range order can be stabilized at finite temperatures, but instead the system will show a cross-over to a regime with local short-range magnetic correlations that mimic the unstable long-range order.

Our numerical simulations indeed corroborate the expected physics. As shown in the summary in Fig.~\ref{fig:1DOverview}(a) the system is devoid of any true phase transition -- the bump in the specific heat exhibits no scaling behavior with increasing system size and the magnetization remains zero for all temperatures. The model does exhibit local, short-range structure, as can be seen in the histograms of the local planar magnetization, $(m_x^{l,p},m_y^{l,p})$, defined on a single triangle (rather than the whole lattice), in Fig.~\ref{fig:1DOverview}(b)-(d). As the temperature is lowered, short-range correlations with a distinctive circular ring structure emerge, reminiscent of the $U(1)$ symmetric structure observed in the global magnetization in the 2D case. The ring in fact has a small $\mathbb{Z}_2$ modulation, most clearly seen in the lowest temperature histogram at $T=0.14$ in Fig.~\ref{fig:1DOverview}(b). The Hamiltonian in this 1D lattice geometry possesses only discrete $\mathbb{Z}_2$ symmetries, in contrast to the $\mathbb{Z}_6$ symmetry intrinsic to the kagome and hyperkagome on account of their $C_3$ lattice rotational symmetry. 

%%%%%%%%%%%%%%%%%%%%%%%%%%%%%%%%%%%%%%%%%%%%
%% Quantum Order-by-Disorder
%%%%%%%%%%%%%%%%%%%%%%%%%%%%%%%%%%%%%%%%%%%%

\section{Quantum Order-by-Disorder} \label{sec:QObD}

Having established the interplay of thermal fluctuations and the emergent $U(1)$ symmetry in the AFM $\GP$-model,
a natural follow-up question is to address the effect of quantum fluctuations, expecting that they will similarly drive
a \emph{quantum} order-by-disorder selection of a discrete set of ground states (not necessarily identical to their thermal
counterparts).

The quantum order-by-disorder calculation shares many similarities with the classical calculation outlined in detail in Sec.~\ref{sec:TObD}.
Therefore, we only briefly summarize the steps of the quantum calculation here and refer the inclined reader to Ref.~\cite{Toth15} for details. For the quantum case, the quadratic fluctuations about the ordered ground state can be captured via a Holstein-Primakoff transformation and subsequent expansion in $1/S$ \cite{Holstein40}. After a Fourier transformation, the resulting quadratic Hamiltonian at order $\mathcal{O}(S)$ in the expansion can be diagonalized, giving the magnon band structure familiar from linear spin-wave theory (LSWT). The quantum correction to the classical ground state energy is then simply the sum of all of the LSWT eigenvalues (as opposed to the sum of the logarithm of the eigenvalues that we saw in the case of the entropy correction for the case of thermal fluctuations). This correction, solely due to quantum zero-point fluctuations, will break the emergent U(1) symmetry of the classical AFM ground-state manifold and is again found to select a discrete six-fold subset of states.

\begin{figure}[t]
    \begin{center}
        \includegraphics[width=\linewidth]{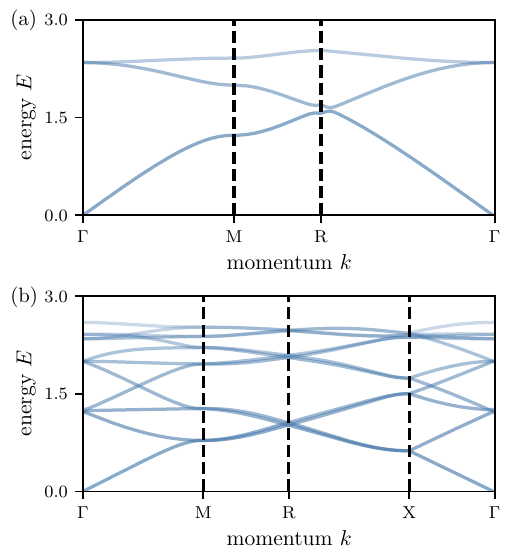}
    \end{center}
    \caption{{\bf Quantum order-by-disorder magnon band structures.}
        Illustrated are the magnon band structures resulting from quantum fluctuations
        for (a) the kagome and (b) the hyperkagome lattice for the AFM ground state corresponding to $\theta=0$.
        The $\Gamma$ point refers to the center of the Brillouin zone,
        $R$ refers to the corner, $M$ refers to the
        center of the edges and $X$ refers to the center of the face.
    }
    \label{fig:QObDBandStructures}
\end{figure}

The magnon band structures from LSWT for a representative example of the AFM ground-state manifold (the case $\theta=0$) are shown
in Fig.~\ref{fig:QObDBandStructures} for both the two-dimensional kagome lattice and the three-dimensional hyperkagome lattice. Note that both band structures exhibit gapless points at zero momentum. This zero energy, zero momentum mode is a reflection of the emergent U(1) symmetry and is an artifact of the approximations used to generate the spectrum. In reality, quantum fluctuations will lift this pseudo-Goldstone mode and fully gap the magnon spectrum, as expected for a model with only discrete Hamiltonian symmetries.

Similar to before we are not interested in the magnitude of the energy correction itself but rather in the difference between states within the ground-state manifold. We find that, for both lattices, this difference can be written as
\be
\Delta E_{Q}(\theta) = \delta_{\rm quantum} \cdot \cos(6\theta) \,,
\ee
with $\delta_{\rm quantum}= 7.9 \cdot 10^{-6}$ for the kagome and $\delta_{\rm quantum}= 7.1 \cdot 10^{-6}$ for the hyperkagome lattice. Exactly as in the case of thermal fluctuations, quantum fluctuations select the six maximally non-coplanar ground states within the AFM ground-state manifold. In other words, both thermal and quantum order-by-disorder drive the selection of the exact same six-fold set of ground states.

%%%%%%%%%%%%%%%%%%%%%%%%%%%%%%%%%%%%%%%%%%%%
%% Discussion
%%%%%%%%%%%%%%%%%%%%%%%%%%%%%%%%%%%%%%%%%%%%

\section{Discussion}
\label{sec:discussion}

We have shown that the classical AFM $\GP$-model, defined on lattices constructed using an elementary 3-site triangle as the building block, exhibits an emergent continuous $U(1)$ symmetry due to a purely accidental ground state degeneracy. This systematic construction allows for a systematic study of the impact of thermal and quantum fluctuations on such an inherently fragile emergent symmetry. Focusing on lattices with a combined spin and lattice $\mathbb{Z}_6$ symmetry, thermal fluctuations generate a rich finite temperature phase diagram with distinct physics depending on the dimensionality of the lattice, as expected due to the renormalization group flow of high-order clock anisotropy in the planar XY model \cite{Jose77,Masaki00}. Interestingly, both thermal and quantum fluctuations select, via the thermal and quantum order-by-disorder mechanisms, the same maximally non-coplanar ground states. It's important to note that not all lattices constructed using the rules discussed in Sec.~\ref{sec:lattices} will possess an underlying $\mathbb{Z}_6$ symmetry. The 2D square-kagome lattice, constructed via connecting corner-sharing triangles, is an obvious example. In this case, the model possesses only discrete $\mathbb{Z}_2$ symmetries, which are highly relevant perturbations that trigger a single finite-temperature transition from the high-temperature paramagnetic phase to a low-temperature $\mathbb{Z}$ Ising ordered phase (see Appendix \ref{appendix:otherlattices} for more details). Let us also note that the order-by-disorder physics of the 2D kagome lattice discussed here complements a number of existing 2D studies, for example, the generic nearest-neighbor models on the kagome lattice discussed in Ref.~\cite{Essafi2017}, and the Kitaev-Heisenberg model on the honeycomb lattice \cite{Price2012,Price2013}. Similarly, the 3D hyperkagome results complement the order-by-disorder phenomena heavily discussed on the pyrochlore lattice, e.g.~in Refs.~\cite{Moessner14,HanYan17,Champion2004,Savary2012,Zhitomirsky2012,McClarty2014}.

The relevance of the model, or more realistically a model with dominant $\GP$ interactions, for materials is not at all clear at this moment in time. Though there are candidate materials in which the $\GP$ interaction appears, it is always sub-dominant and, as far as we are aware, there is no currently known microscopic mechanism to generate a dominant symmetric off-diagonal exchange interaction. However, there is evidence for substantial off-diagonal interactions of $\Gamma$-type in a recent rare-earth based hyperhoneycomb material \cite{Okuma2024}. It has also been suggested that an extended Kitaev model may be of relevance to the rare-earth based kagome lattice compounds of the type $A_2$RE$_3$Sb$_3$O$_{14}$, where $A=\,\,$Mg or Zn and RE is a rare-earth ion \cite{Scheie2016,Dun2016,Dun2017,Paddison2016,Sanders2016}. Future materials studies may uncover more concrete candidate materials.  

Though we have focused on lattices of corner-sharing triangles in this work, the $\GP$-model is perhaps most well-known as a component of the extended Kitaev model on the honeycomb lattice. However, the classical ground states of the model on the honeycomb lattice are actually rather unexciting, for both the FM and AFM cases (see Appendix \ref{appendix:otherlattices} for more details). It exhibits a simple fully FM ground state, with all spins pointing along the $[111]$ direction, for the FM case, and a simple Neel ground state, with the spins on the $A$ sublattice pointing along $[111]$ and on the $B$ sublattice along $[\bar{1}\bar{1}\bar{1}]$ (as well as in both cases a second partner ground state with all spins flipped). The $\GP$-model thus provides an example in which much richer physics can be unlocked by moving beyond honeycomb-based lattices and studying extended Kitaev models on triangular-based lattices.

%%%%%%%%%%%%%%%%%%%%%%%%%%%%%%%%%%%%%%%%%%%%
%% Acknowledgements
%%%%%%%%%%%%%%%%%%%%%%%%%%%%%%%%%%%%%%%%%%%%

\begin{acknowledgments}
    \textit{Acknowledgements.--} 
    We thank 
    	O.~Benton, 
	N.~Perkins,
	U.~Seifert and A.~Rosch 
	for illuminating discussions.
    We acknowledge partial funding from the DFG within Project-ID 277146847, SFB 1238 (project C03). This research was
    supported in part by grant NSF PHY-1748958 to the Kavli Institute for Theoretical Physics (KITP). This work was performed in part at Aspen Center for Physics, which is supported by National Science Foundation grant PHY-2210452.
    The numerical simulations were performed on the Noctua2 cluster at the Paderborn Center for Parallel Computing (PC$^2$). \\
\end{acknowledgments}

{\it Data availability}.--
For all figures shown, the underlying numerical data is available on Zenodo~\cite{zenodo_gamma}.\\

{\it Code availability}.--
The code used to perform the Monte Carlo simulations is made available open-source on Zenodo~\cite{Ramchandani24}.

%%%%%%%%%%%%%%%%%%%%%%%%%%%%%%%%%%%%%%%%%%%%
%% Bibliography
%%%%%%%%%%%%%%%%%%%%%%%%%%%%%%%%%%%%%%%%%%%%
\bibliography{GammaPrime}

%%%%%%%%%%%%%%%%%%%%%%%%%%%%%%%%%%%%%%%%%%%%
%% Appendix
%%%%%%%%%%%%%%%%%%%%%%%%%%%%%%%%%%%%%%%%%%%%
%\clearpage
\appendix

%%%%%%%%%%%%%%%%%%%%%%%%%%%%%%%%%%%%%%%%%%%%
\section{${\bf \GP}$-model on alternative lattice geometries} \label{appendix:otherlattices}
%%%%%%%%%%%%%%%%%%%%%%%%%%%%%%%%%%%%%%%%%%%%
\FloatBarrier

The lattice models discussed in the main text, kagome and hyperkagome, are just two examples of lattices that maintain the emergent $U(1)$ symmetry of the AFM $\GP$-model of a single triangle. Here we briefly discuss the physics of two alternative
lattice geometries, (i) the honeycomb lattice, which
does not exhibit an emergent U(1) symmetry, and (ii)
the square-kagome lattice, which does not possess a $C_3$ lattice rotation symmetry and thus only has a $\mathbb{Z}_2$ symmetry, rather than a $\mathbb{Z}_6$ symmetry.

\begin{figure}[h]
    \centering
    \includegraphics[width=.9\linewidth]{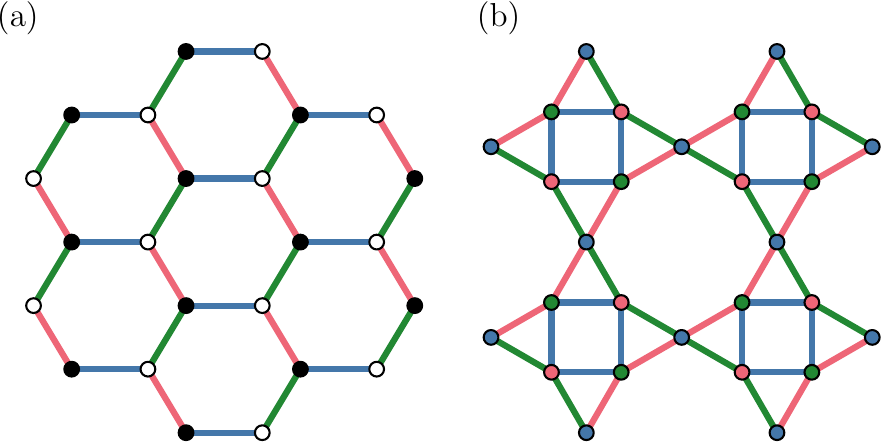}
    \caption{{\bf Alternative lattice geometries.}
        Illustrated are additional lattice geometries for the $\GP$ model.
        These are (a) honeycomb and (b) square-kagome (squagome).}
\end{figure}

%%%%%%%%%%%%%%%%%%%%%%%%%%%%%%%%%%%%%%%%%%%%
\subsubsection*{honeycomb lattice}
%%%%%%%%%%%%%%%%%%%%%%%%%%%%%%%%%%%%%%%%%%%%

The $\GP$ model is most well-known as a component of the extended Kitaev model on the honeycomb lattice, used to describe the physics of the ``Kitaev materials" \cite{Trebst22}. A very natural question then is what is the ground state of the pure $\GP$-model on the honeycomb lattice? The answer turns out to be rather straightforward. For the ferromagnetic model, the ground state is a simple $[111]$ ferromagnet, with the spins on the $A$ and $B$ sublattices given by
\be
\vec{S}_A=\vec{S}_B= \pm \frac{1}{\sqrt{3}}\begin{pmatrix} 1 \\ 1 \\ 1 \end{pmatrix} .
\ee

\noindent
For the anti-ferromagnetic model, the ground state is a simple $[111]$ Neel antiferromagnet, with the spins on the $A$ and $B$ sublattices given by 
\be
\begin{aligned}
    \vec{S}_A & = + \frac{\sigma}{\sqrt{3}}\begin{pmatrix} 1 \\ 1 \\ 1 \end{pmatrix} \,\,,  \,\,
    \vec{S}_B & = - \frac{\sigma}{\sqrt{3}}\begin{pmatrix} 1 \\ 1 \\ 1 \end{pmatrix} \,.
\end{aligned}
\ee
where $\sigma= \pm1$.

%%%%%%%%%%%%%%%%%%%%%%%%%%%%%%%%%%%%%%%%%%%%
\subsubsection*{square-kagome lattice}
%%%%%%%%%%%%%%%%%%%%%%%%%%%%%%%%%%%%%%%%%%%%

The square-kagome (or squagome) lattice is, like the kagome lattice, a two-dimensional lattice of corner-sharing triangles. In other words, it can also be constructed using the lattice
construction rules outlined in the main text that maintain the emergent $U(1)$ symmetry of the AFM $\GP$-model. However, highly relevant for the finite temperature physics of the model, the lattice crucially does not possess a $C_3$ lattice rotation symmetry (unlike the kagome and hyperkagome lattices considered in the main text). Thus, when considering the impact of thermal fluctuations,
the resulting entropy corrections are given by
\be
\Delta S = \delta \cos(2\theta + 2\pi/3) \,, 
\ee
which exhibits a $2\theta$, rather than $6\theta$, dependence. This is in sharp contrast to the kagome and hyperkagome lattices considered in the main text and reflects the distinct underlying symmetry of the $\GP$-Hamiltonian on the square-kagome lattice. 

\begin{figure}
\centering
\includegraphics{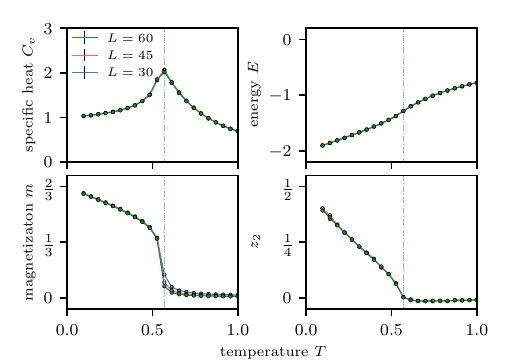}
\caption{{\bf Thermodynamics for the square-kagome lattice.}
Shown are the numerical results from classical Monte Carlo simulations 
for the square-kagome lattice for systems sizes $L=30,45,60$, i.e.~up to a total
of $N=21,600$. The geometry of the lattice 
leads to a distinct finite-temperature phase diagram, resembling the XY model with a $Z_2$, rather than $Z_6$, perturbation. Specifically, the square-kagome lattice exhibits a direct Ising transition to a $\mathbb{Z}_2$ ordered phase, that is captured by a $z_2$ order parameter.
}
\label{fig:squarekagome}
\end{figure}

The physics of the model resembles the planar XY model with a $\mathbb{Z}_2$, rather than $\mathbb{Z}_6$, perturbation. In this case, such a perturbation is relevant even close to the critical point and one expects a single, direct Ising transition from the high-temperature paramagnetic phase to the low-temperature $\mathbb{Z}_2$ ordered phase \cite{Jose77}. This expectation is again confirmed by numerical Monte Carlo simulations, with the results shown in Fig.~\ref{fig:squarekagome}. The specific heat exhibits a single peak at which there is a simultaneous onset of a finite magnetization and $z_2$ parameter, defined as $z_2=\cos(2\theta+2\pi/3)$.

%%%%%%%%%%%%%%%%%%%%%%%%%%%%%%%%%%%%%%%%%%%%
\section{Monte Carlo}
\label{appendix:MonteCarlo}
%%%%%%%%%%%%%%%%%%%%%%%%%%%%%%%%%%%%%%%%%%%%

The technique of choice for the numerical simulations is
Markov Chain Monte Carlo (MCMC).
This technique suits our investigation of the model
since it is an exact technique that allows us to
explore the full thermodynamics including entropic effects
across multiple dimensions. Moreover, it does not suffer from
the sign problem since our model is based on classical Heisenberg spins.

%%%%%%%%%%%%%%%%%%%%%%%%%%%%%%%%%%%%%%%%%%%%
\subsection{Methodology}
%%%%%%%%%%%%%%%%%%%%%%%%%%%%%%%%%%%%%%%%%%%%
The Monte Carlo technique used for these simulations is based on Ref.~\cite{HanYan17}.
Here, the thermalization for the configurations is done in two steps.

\begin{enumerate}
    \item Simulated annealing
    \item Parallel tempering
\end{enumerate}

The measurements are also done using the parallel tempering (replica exchange)
Monte Carlo method.
Each Monte Carlo Sweep (MCS) consists of the following steps.
\begin{enumerate}
    \item Conical local updates based on Ref.~\cite{Alzate19}
    \item Over-relaxation updates
    \item Replica exchange
    \item Measurement
\end{enumerate}

We improved the phase space sampling for the Monte Carlo simulations by using the optimization
discussed in \cite{Alzate19}.
The simulations were implemented in Julia \cite{julia} using the
packages in Ref.~\cite{SpinMC,BinningAnalysis} and the visualizations
are implemented using packages in Ref.~\cite{Makie,MPL}.

%%%%%%%%%%%%%%%%%%%%%%%%%%%%%%%%%%%%%%%%%%%%
\subsection{Parameters for Monte Carlo simulations}
%%%%%%%%%%%%%%%%%%%%%%%%%%%%%%%%%%%%%%%%%%%%

Here we discuss the parameters used for the Monte Carlo simulations.
The number of thermalization sweeps ($N_T$) is done in two steps.
First, a total of $N_T$ thermalization sweeps spread across all
temperatures higher than or equal to the necessary temperature
performed using simulated annealing. Second, $N_T$ thermalization sweeps
performed at the fixed temperature. The number of measurement sweeps
($N_M$) is generally $N_M = 10 N_T$.
Other parameters used when performing the simulations are
given in table $\ref{table:mcParameters}$.

\begin{table}[h!]
    \begin{center}
        \begin{tabular}{l|r}
            \hline\hline
            Parameter                   & Value           \\
            \hline
            Measurement rate            & 1 per sweep     \\
            Replica exchange rate       & 1 per 10 sweeps \\
            Over-relaxation update rate & 1 per sweep     \\
            \hline\hline
        \end{tabular}
    \end{center}
    \caption{{\bf Monte Carlo Parameters.}
        The parameters used for parallel tempering (replica exchange)
        Monte Carlo simulations}
    \label{table:mcParameters}
\end{table}

%%%%%%%%%%%%%%%%%%%%%%%%%%%%%%%%%%%%%%%%%%%%
\subsection{Statistics}
%%%%%%%%%%%%%%%%%%%%%%%%%%%%%%%%%%%%%%%%%%%%
This subsection gives information on the number of measurements
and technique used for the data presented in this work.
For the kagome and hyperkagome lattices, Figure $\ref{fig:MonteCarlo}$ shows data from
parallel tempering Monte Carlo runs.
For each system size, we perform 100 independent runs
with each run containing $10^{7}$ measurements for a total of $10^{9}$
measurements.

Figure $\ref{fig:planarMagnetizationAndCOP}$ shows data from a
Monte Carlo run that records the magnetization vector from
each measurement which is then plotted using a 2D histogram.
The bins are spaced evenly from $-1$ to $1$ with 100 bins in
each direction. The data is gathered from 100 independent
simulated annealing runs with each run containing $10^{6}$ measurements
for a total of $10^8$ measurements.

\end{document}